\documentclass[twocolumn,aps,prl,groupedaddress,showpacs]{revtex4}
\usepackage{epsfig,amssymb}
\def\comment#1{}\def\labell#1{\label{#1}}
\begin{document}
\title{Classical capacity of the lossy bosonic channel: the exact
solution}

\author{V. Giovannetti$^1$, S. Guha$^1$, S. Lloyd$^{1,2}$, L.
  Maccone$^1$, J. H.  Shapiro$^1$, and H. P.
  Yuen$^3$}\affiliation{$^1$Massachusetts Institute of Technology --
  Research Laboratory of Electronics\\$^2$Massachusetts Institute of
  Technology -- Department of Mechanical Engineering\\ 77
  Massachusetts Ave., Cambridge, MA 02139-4307.\\$^3$Northwestern
  University -- Department of Electrical and Computer Engineering,
  2145 N. Sheridan Rd., Evanston, IL 60208-3118.}

\begin{abstract}
  The classical capacity of the lossy bosonic channel is calculated
  exactly.  It is shown that its Holevo information is not
  superadditive, and that a coherent-state encoding achieves
  capacity. The capacity of far-field, free-space optical
  communications is given as an example.
\end{abstract}
 \maketitle

A principal goal of quantum information theory is evaluating the
information capacities of important communication channels.  At
present---despite the many efforts that have been devoted to this
endeavor and the theoretical advances they have produced
{\cite{canali}}---exact capacity results are known for only a handful
of channels.  In this paper we consider the lossy bosonic channel, and
we develop an exact result for its classical capacity $C$, i.e., the
number of bits that it can communicate reliably per channel use. The
lossy bosonic channel consists of a collection of bosonic modes that
lose energy en route from the transmitter to the receiver.  Typical
examples are free space or optical fiber transmission, in which
photons are employed to convey the information.  The classical
capacity of the lossless bosonic channel---whose transmitted states
arrive undisturbed at the receiver---was derived in
{\cite{caves,yuen}}.  When there is loss, however, the received state
is in general different from the transmitted state, and quantum
mechanics requires that there be an accompanying quantum noise source.
In {\cite{sohma}} a first step toward the capacity of such channels
was given by considering only separable encoding procedures.  Here, on
the contrary, it is proven that the optimal encoding is indeed
separable. We obtain the value of $C$ in the presence of loss when the
quantum noise source is in the vacuum state, i.e., when it injects the
minimum amount of noise into the receiver.  Our derivation proceeds by
developing an upper bound for $C$ and then showing that this bound
coincides with the lower bound on $C$ reported in
{\cite{nostro,holevo}}.  Our upper bound results from comparing the
capacity of the lossy channel to that of the lossless channel whose
average \em input\/\rm\ energy matches the average \em output\/\rm\ 
energy constraint for the lossy case {\cite{yuenmax}}. This argument
is analogous to the derivation of the classical capacity of the
erasure channel {\cite{erasure}}.  The lower bound comes from
calculating the Holevo information for appropriately coded
coherent-state inputs. Thus, because the two bounds coincide, we not
only have the capacity of the lossy bosonic channel, but we also know
that capacity can be achieved by transmitting coherent states.
\paragraph*{Classical capacity.--}
 \labell{s:nbwb}
The classical capacity of a channel can be expressed in terms of
the Holevo information
\begin{eqnarray}
\chi(p_j,\sigma_j)\equiv S(\sum_jp_j\sigma_j)-\sum_jp_jS(\sigma_j)
\;\labell{chi},
\end{eqnarray}
where $p_j$ are probabilities, $\sigma_j$ are density operators and
$S(\varrho)\equiv-$Tr$[\varrho\log_2\varrho]$ is the von Neumann
entropy.  Since it is not known if $\chi$ is additive, $C$ must be
calculated by maximizing the Holevo information over successive uses
of the channel, so that $C=\sup_n(C_n/n)$ with
\begin{eqnarray}
C_n=\max_{p_j,\sigma_j}\;\chi(p_j,{\cal
   N}^{\otimes n}[\sigma_j])
\;\labell{defc},
\end{eqnarray}
where the states $\sigma_j$ live in the Hilbert space   ${\cal
H}^{\otimes n}$ of $n$ successive uses of the channel and $\cal N$ is
the completely positive map that describes the channel {\cite{hsw}}.
In our case, $\cal H$ is the Hilbert space associated with the bosonic
modes used in the communication and $\cal N$ is the loss map.  Because
$\cal H$ is infinite dimensional, $C_n$ diverges unless the maximization in
Eq.~(\ref{defc}) is constrained: here we assume
that the mean energy of the input state in each of the $n$
realizations of the channel is a fixed quantity $\cal E$. For
multimode bosonic channels, $\cal N$ is given by
$\bigotimes_k{\cal N}_k$, where ${\cal N}_k$ is the loss map for the $k$th
mode, which can be obtained, tracing away the vacuum noise mode $b_k$,
from the Heisenberg evolution
\begin{eqnarray}
a'_k=\sqrt{\eta_k}\;a_k +\sqrt{1-\eta_k}\;b_k
\;\labell{ch},
\end{eqnarray}
with $a_k$ and $a'_k$ being the annihilation operators of the input and
output modes and $0\le \eta_k \le 1$ is the mode transmissivity (quantum
efficiency).

The main result of this paper is that the capacity of the lossy
bosonic channel, in bits per channel use, is
\begin{eqnarray}
C=\max_{N_k}\sum_k g(\eta_kN_k)
\;\labell{risult},
\end{eqnarray}
where $g(x)\equiv (x+1)\log_2(x+1)-x\log_2x$ and where the maximization is
performed on the modal average photon-number sets $\{N_k\}$ that satisfy the
energy constraint
\begin{eqnarray}
\sum_k\hbar\omega_kN_k={\cal E}
\;\labell{energy},
\end{eqnarray}
($\omega_k$ is the frequency of the $k$th mode). 

We derive Eq.~(\ref{risult}) by giving coincident lower and upper
bounds for $C$.  The right-hand side of Eq.~(\ref{risult}) was shown,
in {\cite{nostro}}, to be a lower bound for $C$ by generalizing the
narrowband analysis of {\cite{holevo}}.  This expression was obtained
from Eq.~(\ref{defc}) by calculating $\chi$ for $n=1$ under the
following encoding: in every mode $k$ we use a mixture of coherent
states $|\mu\rangle_k$ weighted with the Gaussian probability
distribution
\begin{eqnarray}
p_k{(\mu)}=\exp[-|\mu|^2/N_k]/(\pi N_k)\;.\;\labell{pdik}
\end{eqnarray}
This corresponds to feeding the channel the input state
\begin{eqnarray}
\varrho=\bigotimes_k\int d\mu\; p_k(\mu)\;|\mu\rangle_k\langle\mu|
\;\labell{stato},
\end{eqnarray}
which is a thermal state that contains no entanglement or squeezing.
The right-hand side of Eq.~(\ref{risult}) is also an upper bound for $C$.
To see that this is so, let $\bar p_j$, $\bar\sigma_j$ be the optimal
encoding on
$n$ uses of the channel, which gives the capacity $C_n$ of
Eq.~(\ref{defc}). The definition of $\chi$ and the subadditivity of
the von Neumann entropy allow us to write
\begin{eqnarray}
C_n\leqslant S({\cal N}^{\otimes n}[\bar\sigma])
\leqslant\sum_{l=1}^n\sum_kS({\cal N}_k[\varrho_k^{(l)}])
\;\labell{disug},
\end{eqnarray}
where $\bar\sigma\equiv\sum_j\bar p_j\bar\sigma_j$ and   ${\cal
N}_k[\varrho_k^{(l)}]$ is the reduced density operator of the $k$th
mode in the $l$th realization of the channel, which is obtained from
${\cal N}^{\otimes n}[\bar\sigma]$ by tracing over all the other modes
and over the other $n-1$ channel realizations.  The first inequality
in Eq.~(\ref{disug}) comes from bounding $C_n$ by the amount of
information that can be transmitted through a lossless channel with
input state ${\cal N}^{\otimes n}[\bar\sigma]$, viz., the output of
the lossy channel with optimal input state $\bar\sigma$
{\cite{yuenmax}}.  Now let $N_k^{(l)}$ be the average photon number
for the state $\varrho_k^{(l)}$; $\{N_k^{(l)}\}$ must satisfy the
energy constraint (\ref{energy}) for all $l$ {\cite{nota3}}. Moreover,
the loss will leave only $\eta_kN_k^{(l)}$ photons, on average, in the
corresponding output state ${\cal N}_k[\varrho_k^{(l)}]$. This implies
that
\begin{eqnarray}
S({\cal N}_k[\varrho_k^{(l)}])\leqslant g(\eta_kN_k^{(l)})
\;\labell{disug2},
\end{eqnarray}
where the inequality follows from the fact that the term on the right
is the maximum entropy associated with states that have
$\eta_kN_k^{(l)}$ photons on average {\cite{bekenstein,caves}}.
Introducing Eq.~(\ref{disug2}) into (\ref{disug}), we obtain the
desired result
\begin{eqnarray}
C_n\leqslant\sum_{l=1}^n\sum_k g(\eta_kN_k^{(l)})\leqslant
n\max_{N_k}\sum_kg(\eta_kN_k)
\;\labell{disug3},
\end{eqnarray}
where the maximization is performed over the sets $\{N_k\}$ that
satisfy Eq.~(\ref{energy}). Because Eq.~(\ref{disug3}) holds for any
$n$, we conclude that the right-hand side of (\ref{risult}) is indeed
also an upper bound for $C$.

\paragraph*{Discussion.--}\labell{s:disc}
Some important consequences derive from our analysis. First, capacity
is achieved by a single use of the channel ($n=1$) employing random
coding---factorized over the channel modes---on coherent states as
shown in Eq.~(\ref{stato}).  This means that, at least for this
channel, entangled codewords are not necessary and that the Holevo
information is not superadditive.  Notice that the lossy bosonic
channel can accommodate entanglement among successive uses of the
channel, as well as entanglement among different modes in each channel
use.  Surprisingly, neither of these two strategies is necessary to
achieve capacity.  Nor is it necessary to use any non-classical state,
such as a photon number state or a squeezed state, to achieve
capacity; classical (coherent state) light is all that is needed.
Classical light suffices because the loss map $\cal N$ simply
contracts coherent-state codewords in phase space toward the vacuum
state.  Coherent states retain their purity in this process, and hence
the non-positive part of the Holevo information---the second term of
the right-hand side of Eq.~(\ref{chi})---retains its maximum value of
zero.  Despite the preceding properties, quantum effects are relevant
to communication over the lossy bosonic channel.  For example, our
proof does not exclude the possibility of achieving capacity using
quantum encodings, and such encodings may have lower error
probabilities, for finite-length block codes, than those of the
capacity-achieving coherent state encoding.  This is certainly true
for the lossless case.  In particular, it was already known that $C$
can be achieved with a number-state alphabet {\cite{yuen,caves}}; our
work shows that there is also a coherent-state encoding that achieves
capacity for this case. [The two procedures employ the same average
input state, Eq.~(\ref{stato})].  However, the probability of the
receiver confusing any two distinct finite-length number state
codewords is zero in the lossless case, whereas it is positive for all
pairs of finite-length coherent-state codewords.  The lossless case
also provides an example of the possible role of quantum effects at
the receiver: the optimal coherent-state system uses a classical
transmitter, but its detection strategy, can be highly
non-classical~{\cite{hsw}}.  In contrast, the optimal number-state
system for the lossless channel requires a non-classical light source,
but its receiver uses simple modal photon counting.

How well can we approach this capacity using conventional decoding
procedures? Using the coherent-state encoding of Eq.~(\ref{stato})
with either heterodyne or homodyne detection, the amount of
information that can be reliably transmitted is
\begin{eqnarray}
I=\max_{N_k}\sum_k\xi\log_2(1+\eta_kN_k/\xi^2)\labell{hetc}\;,
\end{eqnarray}
where $\xi=1/2$ for homodyne and $\xi=1$ for heterodyne, and where, as
usual, the maximization must be performed under the energy constraint
(\ref{energy}). Equation (\ref{hetc}) has been obtained by summing
over $k$ the Shannon capacities for the appropriate detection
procedure {\cite{caves}}. In general $I<C$: heterodyne or homodyne
detection cannot be used to achieve the capacity.  However, heterodyne
is asymptotically optimal in the limit of large numbers of photons in
all modes, $N_k\to\infty$ for all $k$, because $g(x)/\log_2(x)\to 1$ as
$x\to\infty$.

The capacity expression $C$ can be simplified by using
standard variational techniques to perform the constrained
maximization in Eq.~(\ref{risult}), yielding {\cite{nostro}}
\begin{eqnarray}
C=\sum_k\;g\left(\eta_kN_k(\beta)\right)
\;\labell{risult1},
\end{eqnarray}
where $N_k(\beta)$ is the optimal photon number distribution
\begin{eqnarray}
N_k(\beta)=\frac {1/\eta_k}{e^{\beta\hbar\omega_k/\eta_k}-1}
\;\labell{nkopt},
\end{eqnarray}
with $\beta$ being a Lagrange multiplier that is determined through the
constraint on average transmitted energy.

In the following sections we calculate the capacities of some bosonic
channels. The first two examples help clarify the derivation of
Eq.~(\ref{risult}); the last is a realistic model of
frequency-dependent lossy communication, on which we also evaluate the
performance of homodyne and heterodyne detection.

\paragraph*{Narrowband channel.--}\labell{s:loss}
Consider the narrowband channel in which a single mode of frequency
$\omega$ is employed. In this case, Eq.~(\ref{risult1}) becomes
\begin{eqnarray}
C=\;g\left(\frac{\eta{\cal E}}{\hbar\omega}\right)
\;\labell{nb},
\end{eqnarray}
where $N={\cal E}/(\hbar\omega)$ is the average photon number at the
input. Equation~(\ref{nb}) was conjectured in {\cite{holevo}}, where
it was given as a lower bound on $C$. The following simple argument
shows that $g(\eta N)$ is also an upper bound for $C$.  Consider the
lossless channel that employs $\eta N$ photons on average per channel
use.  Its capacity is given by $\max_\varrho S(\varrho)$, where the
maximization is performed over input states $\varrho$ with mean energy
${\cal E}'=\eta\hbar\omega N$ {\cite{nota}}. The maximum, computed
through variational techniques, is $g(\eta N)$
{\cite{caves,bekenstein}}.  The lossless channel cannot have a lower
capacity than the lossy channel, because both have the same average
received energy, and the set of receiver density operators achievable
over the lossy channel is a proper subset of those achievable in the
lossless system {\cite{yuenmax}}. This implies that $g(\eta N)$ is an
also upper bound on $C$ and hence equal to $C$.

\paragraph*{Frequency-independent loss.--}\labell{s:loss1}
Now consider a broadband channel with uniform transmissivity,
$\eta_k=\eta$, that employs a set of frequencies
$\omega_k=k\;\delta\omega$ for $k\in{\mathbb N}$. In this case,
Eq.~(\ref{risult1}) gives {\cite{nota2}}
\begin{eqnarray}
C=\frac{\sqrt{\eta}}{\ln 2}\sqrt{\frac{\pi {\cal P}}{3\hbar}}{\cal T}
\;\labell{wb},
\end{eqnarray}
where ${\cal T}=2\pi/\delta\omega$ is the transmission time, and
${\cal P}={\cal E}/{\cal T}$ is the average transmitted power.
Equation~(\ref{wb}) was derived for the lossless case ($\eta=1$) in
{\cite{yuen}} and was shown to provide a lower bound on $C$ in
{\cite{nostro}}. In order to show that the right-hand side of
Eq.~(\ref{wb}) is also an upper bound, consider the lossless broadband
channel in which the average \em input\/\rm\ power is equal to
$\eta {\cal P}$, viz., the average \em output\/\rm\ power of the lossy
channel. According to {\cite{yuen}}, the capacity of this channel is
$(\sqrt{\pi\eta {\cal P}/3}){\cal T}/\ln 2$, which coincides with the
right-hand side of Eq.~(\ref{wb}).  The
reasoning given above for the single-mode case now
implies that the broadband lossless channel's capacity cannot be less than
that of the broadband lossy channel, thus completing the proof.


\begin{figure}[hbt]
\begin{center}
\epsfxsize=.6
\hsize\leavevmode\epsffile{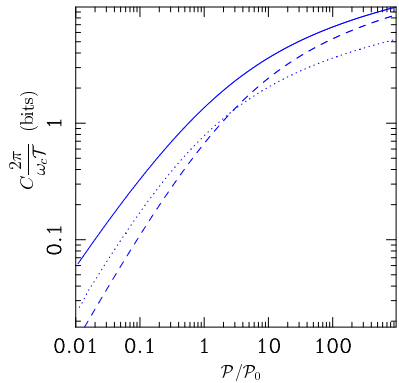}
\end{center}
\caption{Capacities of the far-field free-space optical channel as a
  function of the input power $\cal P$ (in the plot ${\cal P}_0\equiv
  2\pi\hbar c^2L^2/(A_tA_r)$). The solid curve is the capacity $C$
  from Eq.~(\ref{ffc}), the other two curves are the capacities $I$
  from Eq.~(\ref{chethom}) achievable with coherent states and
  heterodyne detection (dashed curve) or coherent states and homodyne
  detection (dotted curve). Note that the heterodyne detection $I$
  approaches the optimal capacity $C$ in the high-power limit.}
\labell{f:num}\end{figure}

\paragraph*{Far-field, free-space optical communication.--}\labell{s:farfield}
Consider the free-space optical communication channel in which the
transmitter and the receiver communicate through circular apertures of
areas $A_t$ and $A_r$ that are separated by an $L$-m-long propagation
path.  At frequency $\omega$ there will only be a single spatial mode
in the transmitter aperture that couples appreciable power to the
receiver aperture when the Fresnel number $D(\omega) \equiv
A_tA_r(\omega/2\pi cL)^2$ satisfies $D(\omega) \ll 1$, \cite{ff}. This
is the far-field power transfer regime at frequency $\omega$, and
$D(\omega)$ is the transmissivity achieved by the optimal spatial
mode.  A broadband far-field channel results when the transmitter and
receiver use the optimal spatial modes at frequencies up to a critical
frequency $\omega_c$, with $D(\omega_c)\ll 1$.  In this case we use
$\eta_k=D(\omega_k)$ in Eq.~(\ref{risult1}), and the capacity $C$
becomes {\cite{nota2}}
\begin{eqnarray}
C=\frac{\omega_c{\cal T}}{2\pi
y_0}\int_0^{y_0}dx\;g\!\left(\frac
   1{e^{1/x}-1}\right)
\;\labell{ffc},
\end{eqnarray}
where $y_0$ is a dimensionless parameter inversely proportional to the
Lagrange multiplier $\beta$, which is determined from the power
constraint
\begin{eqnarray}
{\cal P}=\frac{2\pi\hbar c^2L^2}{A_tA_r}\int_0^{y_0} \frac{dx}x\frac
1{e^{1/x}-1}
\;\labell{e1nerg}.
\end{eqnarray}
Although $C$ is proportional to the maximum frequency $\omega_c$,
this factor cannot be increased without bound, for fixed transmitter and
receiver apertures, because of the far-field assumption.
Figure~\ref{f:num} plots $C$ versus $\cal P$ obtained from numerical
evaluation of Eqs.~(\ref{ffc}) and (\ref{e1nerg}).

\begin{figure}[hbt]
\begin{center}
\epsfxsize=.6\hsize\leavevmode\epsffile{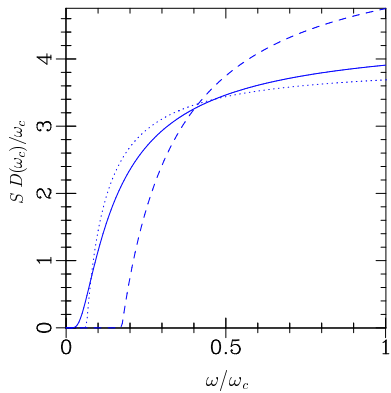}
\end{center}
\caption{Power spectrum $S\equiv\omega_kN_k$ for the far-field
  free-space channel plotted versus frequency in the continuum regime
  {\cite{nota2}}. The solid curve is for optimal capacity, the dotted
  curve is for homodyne detection, and the dashed curve is for
  heterodyne detection. Here ${\cal P}/{\cal P}_0=3$. In contrast to
  the frequency-independent lossy channel, all of these coherent-state
  encodings preferentially employ high frequencies instead of low
  frequencies.  This marked change in spectral shaping is due to the
  transmissivity's having a quadratic dependence on $\omega$. }
\labell{f:enne}\end{figure}

To compare the capacity of Eq.~(\ref{ffc}) with the information
transmitted using heterodyne or homodyne detection, we perform the
Eq.~(\ref{hetc}) maximization. The Lagrange multiplier technique gives
the optimal value   $N_k(\beta)=\max\left\{
1/({\beta\hbar\omega_k})-{\xi^2}/{\eta_k}\;,\;0\right\}$, plotted in
Fig.~\ref{f:enne}.  [Notice that the non-negativity of this solution
forbids the use of frequencies lower than
$\omega_{0}\equiv\xi^2\beta\hbar\omega^2_c/D(\omega_c)$.] With this
photon number distribution, Eq.~(\ref{hetc}) becomes
\begin{eqnarray}
I=\xi{\omega_c{\cal T}}\left(1/{y_0}-1+\ln y_0\right)/({2\pi\ln 2})
\;\labell{chethom},
\end{eqnarray}
where $y_0$ is now determined from the condition   ${\cal
P}=\xi^2{2\pi\hbar c^2L^2} (y_0-1-\ln y_0)/({A_rA_s})$.  We have
plotted $I$ versus $P$ in Fig.~\ref{f:num} for heterodyne and homodyne
detection.  At low power, the noise advantage of homodyne makes its
capacity higher than that of heterodyne.  At high power levels
heterodyne prevails thanks to its bandwidth advantage, and its
capacity approaches $C$ asymptotically.

\paragraph*{Conclusions.--}\labell{s:co}
We have derived the classical capacity of the lossy multimode bosonic
channel { when the average energy devoted to the transmission is
  bounded}. Interestingly, quantum features of the signals (such as
entanglement or squeezing) are not required to achieve capacity,
because an optimal coherent-state encoding exists. At the decoding
stage, however, quantum effects might still be necessary (e.g., in the
form of joint measurements on the output) as standard homodyne and
heterodyne measurements are not optimal, except for the high power
regime where heterodyne detection is asymptotically optimal.  The
focus of this paper has been the lossy channel with minimal
(vacuum-state) noise. A more general treatment would include
non-vacuum noise, and would allow for amplification.

This work was funded by the ARDA, NRO, NSF, and by ARO under a MURI
program.


\begin{references}
\bibitem{nota3} The concavity of $g(x)$ implies that this constraint
   is not strictly necessary: it suffices to require that the average
   over $l$ be fixed, i.e.       $\sum_{k,l}\hbar\omega_kN_k^{(l)}/n={\cal
E}$.
\bibitem{nota} In the noiseless case the maximization of the Holevo
   quantity (\ref{defc}) yields the von Neumann entropy of the input
   state, which is a subadditive quantity.
\bibitem{nota2} Notice that in the high-power regime, the sums in
   Eqs.~(\ref{energy}) and (\ref{risult1}) can be replaced with
   integrals.
\end{references}
\end{document}